# Improving Solar and PV Power Prediction with Ensemble Methods

L. A. Dao*, L. Ferrarini*, D. La Carrubba*

*Dipartimento di Elettronica, Informazione e Bioingegneria,
Politecnico di Milano, P.za L. da Vinci 32, 20133 Milano (Italy)
e-mail: luca.ferrarini@polimi.it

Abstract: Estimation of the generated power of renewable energy resources is in general important for planning operations as well as demand balance and power quality. This paper addresses the problem of the estimation of the short-term (3-hour ahead) and medium-term (1-day ahead) generated power of a photovoltaic plant. Firstly, the design of day-ahead solar radiation predictors is investigated with different setups of time series models, and with their combinations with the weather forecast services using ensemble methods. Support Vector Machine methods are also adopted in this stage, to cluster data. Secondly, under a similar ensemble framework, the generated power prediction is investigated. The whole generated power and solar radiation prediction tasks are then implemented on a low-cost, embedded mini PC module Raspberry Pi 3. As an application, the prediction is employed in the control system of a typical microgrid settings focusing on energy management problem. The impact of the quality of generated power prediction on the performance of the controller is also evaluated in this paper.

*Keywords:* Ensemble prediction methods, Generated power prediction, Renewable energy sources, Microgrids.

## 1. INTRODUCTION

Driven by environmental concerns and the fact that fossil fuel price is volatile, the employment of renewable energy resources is getting more and more attention all over the world. Indeed, the global renewable power capacity more than doubled during the last decade from approximately 1000 gigawatts in 2007 to 2195 gigawatts in 2017; in 2017 alone, 70% of the net addition of global power capacity was from renewable energy which marked the incredibly increasing trend of renewable energy (Hales, 2018).

Notwithstanding the promising advantages it may bring, the penetration of renewable energy sources also puts many challenges to the power system. One disadvantage of renewable energy is its variability, thus the reliability of its supply; furthermore, the difficulty in predicting their outputs due to high correlation to the weather (and the weather is difficulty to predict) may cause significant problems of demand balance and power quality to the distribution network and the grid in general if large capacity renewable energy sources are connected.

To deal with these issues, one of the first, clear directions is to improve the quality of renewable energy production prediction and secondly to compensate the prediction errors through adequate control and management techniques. The present paper is in this line. Other research lines are based on the involvement of the demand side to play a more crucial role in supporting the power system, the exploitation of energy storage system, the design of more autonomous microgrids (L. A. Dao, Piroddi, & Ferrarini, 2015)

Regarding the quality of renewable energy production prediction, in the literature, two main approaches have been studied. The first one is based on the use of single prediction models for Solar Radiation (SR) or PV-related generation plants, adopting a wide range of standard techniques including Artificial Neural Networks (ANN) (Sulaiman, Rahman, & Musirin, 2009), Extreme Learning Machine (EML) (Hossain, Mekhilef, Danesh, Olatomiwa, & Shamshirband, 2018), Support Vector Machine (SVM) (Chen, Li, & Wu, 2013), Autoregressive Integrated Moving Average model (ARIMA) (Yang, Jirutitijaroen, & Walsh, 2012), and so on. A set of more complex predictors are investigated through combining of multiple different prediction models and weather forecasting services. They are called ensemble prediction methods. The combination is expected to inherit the best characteristic from each individual prediction model, and compensate for the single predictor's errors. Paper (Zhou, 2012) suggests that besides the accuracy of the single predictors, their diversity is a crucial element to obtain a good ensemble. Some recent works have shown increasing attention on the use of ensemble prediction methods such as (Chakraborty, Marwah, Arlitt, & Ramakrishnan, 2012), which combines Motif based prediction, k-NN prediction and naïve Bayes prediction, or (Ji & Chee, 2011), which employs Autoregressive and Moving Average (ARMA) models together with Time Delay Neural Networks (TDNN).

The present paper investigates different techniques to perform SR and Generated Power (GP) prediction using different ensemble methods, integrating also meteorological prediction services. In addition, an implementation is discussed on a low-cost, embedded mini PC module Raspberry Pi 3 model B which autonomously performs a comprehensive set of tasks, ranging from collecting the meteorological data and meteorological prediction data from meteorological service providers to data cleaning, single model development and adaptation, ensemble prediction.

This paper largely extends (Le Anh Dao, Piroddi, & Ferrarini, 2017), where some preliminary ideas on ensemble method were investigated. Firstly, in this paper, a new approach is proposed with the use of hierarchical structure in the general scheme of SR prediction whereas a machine learning technique is adopted to cluster the overall set of data into several subsets before dealing with the ensemble method at the second layer of the structure. Furthermore, the investigation is studied on much comprehensive weather data set, different sampling time and horizons. An implementation on Raspberry Pi 3 with Python programming language and Linux operating system is also provided in this paper, its application and numerical results in a smart grid scenario are also computed and described. Specifically, the smart grid scenario here considers an energy management optimization of a microgrid including smart buildings, PV power production facilities and an energy storage unit. The control system is developed based on Model Predictive Control (MPC) approach which is well-suited to deal with a large amount of constraints and multiple objectives that have to be imposed in our control problem (Morari & H. Lee, 1999); the exploitation of this approach in the context of smart grid has been widely discussed in the literature (L. A. Dao, Dehghani-Pilehvarani, Markou, & Ferrarini, 2019).

The paper is organized as follows. In Section 2, the considered case study is described along with some basic analysis of the data available, while in section 3 the selected ensemble technique is reviewed and models described. Section 4 describes the implementation on Raspberry Pi, and the obtained results. Finally, Section 5 concludes the paper.

## 2. DESCRIPTION OF THE CASE STUDY

PV plants and meteorological stations analysed in this paper are located in the North of Italy. The overall goal is to develop a reliable PV GP prediction for a pilot plant, starting from its historical data, the SR historical data and the meteorological forecast data from a service provider for the same location. The illustration of these periods is depicted in the following figure:

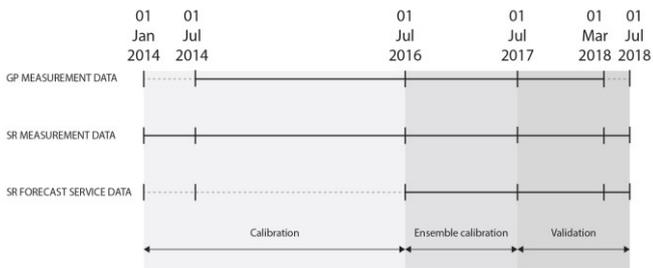

Fig. 1. Periods of available data of GP (1-min and 5-min sampling time), SR (10-min sampling time) and SR meteo-rological forecast services (3-hour sampling time, dash lines for unavailable data)

Considering the common periods between them, the available data of GP, SR and SR forecast data from the meteorological forecast service are separated into three periods:
- Calibration: to train the time series models for SR and GP predictions (period 1)
- Ensemble calibration: to train ensemble model for GP and SR (period 2)
- Validation: to validate the proposed predictors. As the GP data is available only till 01-March-2018, the validation period for GP is ended on 01-March-2018 (period 3).

Concerning data pre-processing, detection production outliers step is performed by firstly comparing an original production datum (i.e., 1-min sampling time) with the average of its neighbors and then also by matching then-resampled-to-10-min-sampling production data with the original SR data.

While most of the available data points present high correlation between SR and GP data, there is still a significant number of points that do not match this behavior: they are either with high GP and zero SR or low production with high SR. They are all considered as outliers. The removed samples are regarded as missing data. Isolated missing data (up to sequences of three missing samples) are substituted with the average value of their two closest available data. Finally, to match the typical sampling rate of energy management system in the microgrid, which will be discussed in Section 4, the data are resampled to 15-minute sampling time.

## 3. ENSEMBLING METHOD

### 3.1 General framework

The general framework here considered for our GP prediction is based on a two-step prediction, i.e., SR and GP predictions. Regarding SR prediction, at first, a superior level Support Vector Machine algorithm (SVM) is designed to classify the day with better or worse meteorological forecast service with respect to the time series model (ARI). This consideration will separate the SR data into two subsets; and as following two ensemble blocks ES1 and ES2 are designed to be applied to the two different subsets of the data (see Fig. 2). Each of the ensemble blocks (ES1 and ES2) contains two groups of predictors: the first group is obtained from the meteorological forecasting service which is updated every 6 hours, whereas the second group is based on models constructed using SR historical data which is performed every 15 minutes. The outputs of all predictors are combined in an ensemble predictor model (i.e., ES1 and ES2). As a final stage, the outputs of all ES1 and ES2 are combined to form the final SR prediction which is input of the GP prediction.

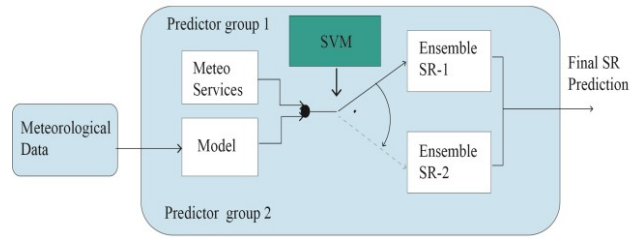

Fig. 2. SR ensemble method with SVM.

To classify the SR data, we here employed standard SVM classifier with a Gaussian kernel as discussed in (Schölkopf & Smola, 2002). The training data set is defined as follows:

$$\{x_i, y_i\}, i = 1, 2, \dots, N \qquad (1)$$

where $N$ is the number of samples, $x_i$ are vector of different inputs and $y_i$ is class label of corresponding outputs. In our consideration, the inputs include the following types:

- The change of the day-ahead SR forecast (taken from meteorological service) with respect to SR measurement data in the previous day;
- Time instant of a day (i.e., the first, second, …, 96th instants of a day).

The consideration of the first input type comes from the observation that ARI model would not be able to catch the large variation of the SR measurement data from one day to another since this model employs only historical SR data in its model. On the other hand, in general, the forecast service knows quite well if the big change of SR is about to happen. Regarding the second input type, the meteorological service data would get out-of-date with respect to the ARI model which is updated every 15minutes instead of 6 hours of the meteorological service.

On the other hand, the output $y_i$ are labeled as either "Good" or "Bad". "Good" label implies a better prediction is obtained from the meteorological service with respect to ARI model and vice versa for "Bad" label. The SVM, ES1 and ES2 are all trained in period 2 (i.e., ensemble calibration period) and then validated in period 3 (i.e., validation period). All the forecasts are obtained by MATLAB SVM toolbox. Then, at any time instant, if $y_i$ is labeled as "Good", then the parameter of ES1 is employed to combine the two SR predictions (i.e., meteorological forecast service and ARI model) and similarly, ES2 used in the case of $y_i$ is labeled as "Bad".

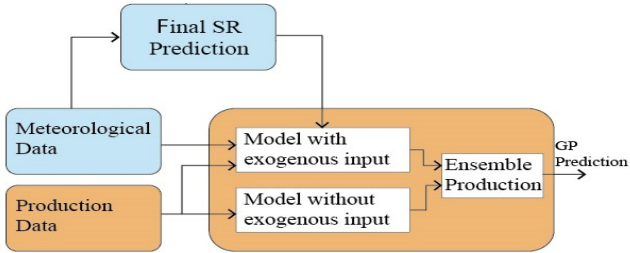

Fig. 3. General scheme for GP prediction using ensembles

Regarding GP prediction, similarly, two different groups of predictors are employed, one based on historical GP data and meteorological measurements, while the second one is only based on GP data (see Fig. 3). The outputs of all predictors are combined in an ensemble predictor, to provide the final GP prediction.

### 3.2 Solar radiation time series models

For SR prediction purposes, the ARI (*p, d*) (autoregressive integrated) class model has been selected which is a special case of the general ARIMAX (*p, d, q, m*) model with two values of *q* and m are set to zero. The description of ARIMAX model can be find in (Williams, 2001).

In order to evaluate the model, ACF (Autocorrelation Function) and PACF (Partial Autocorrelation Function) are employed. Based on this analysis, the decision to drop the MA part from the prediction models is derived and the following candidate models are considered in the ensemble method:

- ARI model with d = 96, p= [1, 2, 3, 4, 94, 95, 96, 97]
- ARI model with d = 96, p= [1, 2, 3, 4]

- The prediction obtained from the meteorological forecast service

where the I part of ARI models performs a 24-hour differencing of the original data, given that the data are sampled every 15 minutes; since the 1st, 2nd, 3rd, 4th, 94th, 95th, 96th and 97th samples of the PACF are statistically different from 0. Shorter ARI models, with $p = [1, 2, 3, 4]$ have also been employed to allow the models to capture different dynamic characteristics of the data and thus enriching the diversity of the ensembled predictors.

### 3.3 PV production time series model

Regarding the GP prediction, both ARI (*p, d*) and ARIX (*p, d, m*) (autoregressive integrated with exogenous inputs) models have been developed. The ARI model relies only on measured production data. On the other hand, the ARIX model employs the estimated SR as exogenous input, which provides additional information for the estimation of the GP. Even if in the literature, the temperature has often been used together with the SR to model the GP (Accetta, Piroddi, & Ferrarini, 2012), here the addition of the temperature is not particularly useful.

Similar to the SR data, the following single models are considered in the ensemble method:

- ARI model: *p* = [1, 2, 3, 4, 94, 95, 96, 97], *d* = 96;
- ARIX model: *p* = [1, 2, 3, 4, 94, 95, 96, 97], *d* = 96, *q* = [1, 2, …, 12, 94, 95, 96, 97]).

### 3.4 Ensemble predictions

The considered ensemble method aggregates multiple predictors via a linear combination of the outputs. The corresponding weights are obtained by minimizing the distance between the ensemble predictor output and the measured data as follows:

$$\min J(\theta) = \sum_{t=1}^{k} \left\| P(t) - \hat{P}(t) \right\|_2^2 \qquad (2)$$

where $\hat{P}(t)$ is a linear combination of different predictors (two predictors for each of Ensemble SR-1, Ensemble SR-2 and Ensemble Production); assuming that measured data $P(t)$ and output of individual predictors $P_i(t)$ are available in a time inteval from 1 to *k*.

In the SR prediction, the combination is depicted as in Fig. 4. Two alternative settings of the ensemble method are considered here. In the first one, the predictions obtained with various models are first classified into groups, depending on the prediction horizon (4, 8, 12, 24, 36, 48, 60, 72, 84 and 96 steps ahead prediction or 1, 2, 3, 6, 9, 12, 15, 18, 21 and 24 hours ahead). In the second setting, denoted by basic ensemble framework, the predictions are not grouped with respect to the prediction horizons.

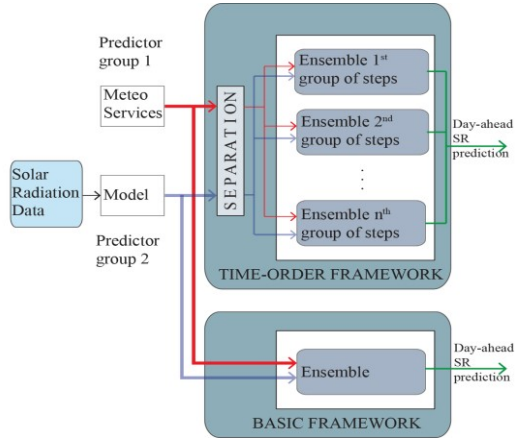

Fig. 4. Ensemble framework in SR prediction

The structure of the ensemble model for GP prediction is similar to the previously discussed one for SR prediction (see Fig. 4). However, due to very similar performance between GP prediction using time-order and basic frameworks for ensemble method we will report only the result of a simpler one (i.e., basic framework).

## 4. IMPLEMENTATION

The implementation of the GP and SR prediction is developed on a Raspberry Pi with Python programming language and Linux operating system. The software architecture of the implementation is shown in Fig. 5, where the blue box represents the Raspberry Pi which contains various functions developed using Python (gray blocks) together with the database tables based on SQLite database storage system (light blue blocks). On the other hand, blocks outside of the Raspberry Pi are servers in which production, meteorological data, meteorological forecast service data, and GP and SR prediction data are stored.

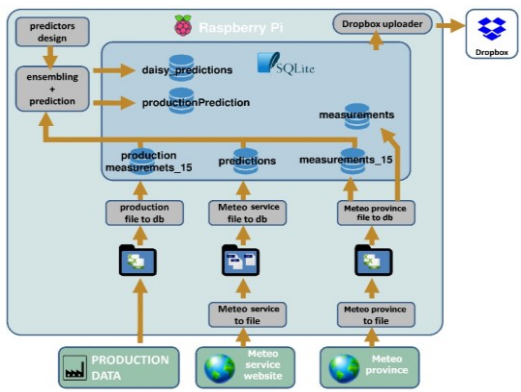

Fig. 5. GP prediction system software architecture

Generally, the prediction system is separated into two main phases. In the offline phase, the historical data from different servers are downloaded then saved into different data tables for both raw and processed data. The processed data is then also employed to derive the system model order and parameters which are saved in the predictor design block. On the other hand, in the online phase which is run every 15 minutes, new GP, SR and meteorological data are searched and put into the corresponding database tables if available. These online steps are performed one minute before the clock minute time of 00, 15, 30 and 45 so that all the necessary data are available for the prediction tasks to run. The results of SR and GP prediction are then saved in the device and uploaded into a cloud storage service – Dropbox.

### 4.1 Solar radiation results

Regarding SR prediction, apart from collecting SR predictions from the available meteorological forecast service, various models for prediction using different model classes and the combinations thereof obtained with the ensemble method have been tested. The obtained models have been compared using a standard Root Mean Square Error (RMSE) criterion:

$$RMSE(t) = \frac{100\%}{M} \sqrt{\frac{\sum_{j=1}^{j=h}\left(\hat{P}(t,j) - P(t,j)\right)^2}{h}} \quad (3)$$

where $M$ is the peak values of SR and GP in the studied measurement data (introduced for normalization purposes). At every time instant, SR prediction or GP prediction is performed and the corresponding value of RMSE [%] is collected. The comparison is then based on the average values of the RMSE [%] for each predictor over the validation data.

The results of the ensemble methods together with naïve predictor and the forecast service for SR prediction are summarized in Table 1 which contains the values of $RMSE_m$ and $RMSE_s$ for the results of medium-term prediction (one-day ahead) and short-term prediction (3-hour ahead), respectively. Although various combinations of the single models have been tested, only the best ones are reported here.

Table 1. List of considered predictors for SR and GP prediction

| Predictor ID | Method | $RMSE_s$ | $RMSE_m$ |
|---|---|---|---|
| SOLAR RADIATION | | | |
| SR-1 | Persistent predictor | 10.8 | 13.9 |
| SR-2 | Forecast service | 7.7 | 10.8 |
| SR-3 | Ensemble method (Basic-order framework) | 4.8 | 7.9 |
| SR-4 | Ensemble method (Time-order framework) | 4.6 | 7.9 |
| SR-5 | Ensemble method (Time-order framework) with SVM | 4.5 | 7.7 |
| GENERATED POWER | | | |
| GP-1 | Persistent predictor | 6.1 | 9.4 |
| GP-2 | ARIX model using SR2 as exogenous input | 6.0 | 9.4 |
| GP-3 | ARIX model using SR4 as exogenous input | 4.9 | 8.0 |
| GP-4 | Ensemble method | 4.6 | 7.7 |

As for one-day ahead SR prediction, the prediction obtained with the meteorological service provides a significant improvement with respect to the one obtained from the naïve predictor. This result is reasonable as the naïve predictor relies mostly on the SR of the previous day while the service typically employed advance prediction techniques and rich data sources including even satellite pictures and very long-term historical data source. Beside this prediction, some prediction model based on ARI and AR techniques is designed to catch some dynamics in the SR data. These prediction models are expected to supplement the final ensemble model the characteristic of the SR data which are not considered in the meteorological service. In fact, the models using ensemble methods (SR-3, SR-4 and SR-5) have 2.9 points of RMSE less than the meteorological service (SR-2) which accounts for a 26.8% improvement in the medium term.

As for the short-term prediction (3 hours), SR-3 performs even much better than the meteorological forecast service since it exploits the most recent measurement data which are updated every 15-minutes, while the meteorological forecast service is updated in a longer sampling time of 6-hour. A slightly improvement is observed by using ensemble method with time-order framework respect to the ensemble method with basic framework. In the end, about 40.3% improvement is recorded for ensemble method SR-4 with respect to the meteorological forecast service. In both short and medium-term predictions, the combination between SVM and ensemble method provides slightly better performance than the best one obtained among other predictors.

*4.2 GP prediction results*

As already discussed, the evaluation of GP prediction is discussed on 8 months of validation data from 01 July 2017 to 28 February 2018. The GP results are reported in the same table of SR prediction results (Table 1).

At first, the impact of quality of SR prediction on GP prediction can be assessed through the comparison between GP-2 and GP-3. Indeed, the inability to predict the SR with a good enough level causes a significant accuracy loss also in the GP prediction. Table 1 also shows that using the SR prediction from the meteorological service does not provide any better results than a very simple predictor – the naïve predictor GP-1. On the contrary, the GP predictor using the best-obtained SR prediction (SR-5) provides significant improvement for GP prediction: GP3 model improves to 8.0 (for one-day ahead horizon) and 4.9 (3-hour ahead horizon) of prediction RMSE of GP2. Some combinations of different predictors can achieve a further improvement reducing RMSE down to 4.6 (for 3-hour horizon) and 7.7 (for one-day horizon) with respect to the best one obtained from single model GP-3. Notice that different combinations of single models have been tested, but only the results of GP (i.e., a combination of all the mentioned predictor in section of PV production prediction of this chapter) are reported as GP-4 provides the best results out of all combinations. In the end, this combination GP-4 provides significantly better results than GP-2 or standard GP predictor. By applying GP-4, the prediction errors of the standard GP predictor reduce from 6.1 (for 3-hour horizon) and 9.4 (for one-day horizon) to 4.6 (for 3-hour horizon) and 7.7 (for one-day horizon) respectively which accounts for 24.6% and 18% improvements, respectively.

## 4. IMPACT OF DAY-AHEAD GP PREDICTION

In this section the impact of the different GP prediction is evaluated on control performance to operate a typical microgrid. The considered microgrid contains energy a storage, flexible loads, PV plants, and a point of common coupling with the national electricity grid as shown in Fig. 6. Briefly, the control system aims to maximize the benefit coming from trading electricity with the market together with minimizing the monetary penalties (or imbalance charge) imposed to the microgrid due to violation of the promised power with the market in the day-ahead market. The interested readers are invited to read (L. A. Dao et al., 2015) for further details on the microgrid settings as well as the development of employed MPC algorithm. The GP prediction are clearly used in the development of the control system, based on MPC.

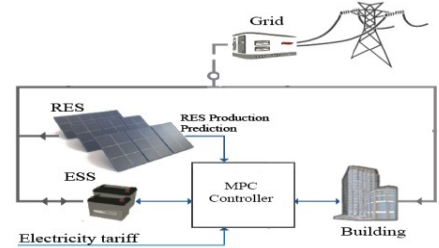

Fig. 6. Typical microgrid settings and MPC controller

Applying the same development of MPC algorithm in (L. A. Dao et al., 2015) to the considered microgrid, we obtain different performances based on different predictors. In this work, the performance of the control system is analyzed in 4 different cases:

- SMPC-P: perfect day-ahead PV GP prediction – RMSE = 0% is considered (the future PV GP power is assumed to be known over the whole day-ahead or day-ahead prediction horizon);
- SMPC-2: day-ahead PV GP prediction is done with the real ensemble predictor discussed in previous section with RMSE of 10.08% in the considered day.
- SMPC-1, SMPC-3 and SMPC-4: synthetic day-ahead PV GP prediction data with RMSE of 5.19%, 14.85% and 19.65% are considered correspondingly.

To perform the analysis considered in this section, we are here adopting the same consideration in (Teleke, Baran, Bhattacharya, & Huang, 2010) to emulate the day-ahead power production (with respect to which the imbalance charges are calculated) by artificially adding a white Gaussian noise to the exact production profile. This noise models the prediction error that would be obtained in practice. Different values of SNR (Signal-to-noise ratio) are set to model the above prediction errors. As a result, the day-ahead predictions in four cases (SMPC-1, SMPC-2; SMPC-3; SMPC-4) are shown in Fig. 7.

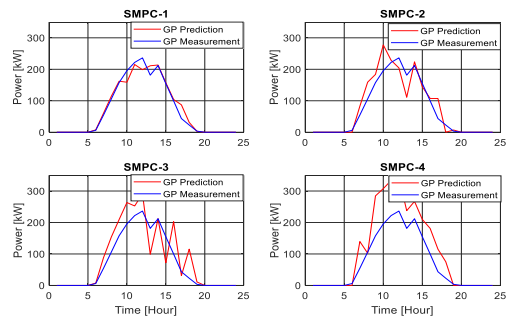

Fig. 7. Day-ahead prediction errors in four cases

Apparently, as shown in Fig. 8, the PV GP prediction error determines a significant performance reduction. An amount of €91.0 for imbalance charge is cut from the total benefit of the microgrid due to a 19.65% of prediction error in SMPC-4. Fig. 8 shows the quality of reference power tracking or PCC power tracking in case SMPC-4 and the corresponding behavior of the ESS. The ESS behavior suggests that the prediction error in SMPC-4 is too big for such an ESS (i.e., Power: 100 kW, Energy level: 150 to 500 kWh) to handle, e.g., from 13 o'clock to 18 o'clock the ESS needs to discharge in order to help the microgrid track the reference power but the ESS energy already reaches the lower limit. As for imbalance charge, much lower penalties are observed in the case of SMPC-1, SMPC-2, SMPC-3 and especially very small penalty in the case of perfect GP prediction in SMPC-P.

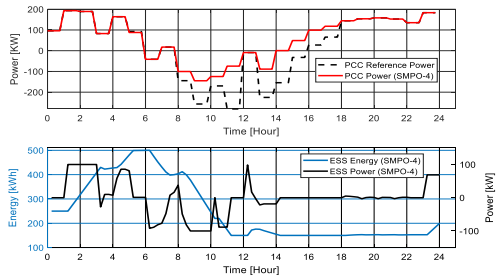

Fig. 8. PCC Power tracking, ESS energy and power (SMPC-4)

At the end, comparing the total benefits in the above tests, we can compute the percentage decrease of total benefit with respect to the ideal perfect prediction case as a function of RMSE of GP prediction: we pass from 1% of decrease of total benefit with RMSE of about 5% (SMC-1) to 26% of decrease of total benefit with RMSE of about 20% (SMPC-4). These observations imply that the GP prediction accuracy can have a great impact on the overall system performance.

**Table 2. Simulation results**

| Case ID | Trading [€] | Charge [€] | benefit [€] |
|---|---|---|---|
| SMPC-P | 379.2 | 2.3 | 377.0 |
| SMPC-1 | 376.8 | 4.8 | 372.0 |
| SMPC-2 | 371.2 | 27.4 | 343.7 |
| SMPC-3 | 372.1 | 44.2 | 328.0 |
| SMPC-4 | 367.9 | 91.0 | 276.9 |

\* SMPC-2 uses GP prediction designed in previous chapter

## 5. CONCLUDING REMARKS

The paper investigates different predictors for SR and GP in a pilot PV plant including ensemble methods for combining multiple predictors. This combination provides significantly more accurate results compared to the one obtained from a commercial meteorological forecast service (for SR prediction) and standard predictor for GP prediction (using ARIX model with SR prediction from the service is exogenous input). In particular, with the analyzed data set, an improvement of approximately 26% (in the medium term) to 40% (short term) is achieved for SR. As for GP prediction, the analysis shows that the dependence of the quality of the GP prediction on the quality of SR prediction might be significant. The ensemble GP prediction here proposed improves approximately 18% (medium term) to 25% (in the short term) with respect to the naïve predictor. Future research lines include the integration with more sophisticated clustering techniques (different ensemble for different clusters), the exploitation of other meteorological data like cloud coverage and wind speed, the correlation with predictions made in neighbouring sites.

REFERENCES

Accetta, G., Piroddi, L., & Ferrarini, L. (2012). Energy production estimation of a photovoltaic system with temperature-dependent coefficients. *2012 IEEE Third International Conference on Sustainable Energy Technologies (ICSET)*, 189–195.

Chakraborty, P., Marwah, M., Arlitt, M., & Ramakrishnan, N. (2012). Fine-grained photovoltaic output prediction using a Bayesian ensemble. *Proceedings of the National Conference on Artificial Intelligence*, *1*, 274–280.

Chen, J. L., Li, G. S., & Wu, S. J. (2013). Assessing the potential of support vector machine for estimating daily solar radiation using sunshine duration. *Energy Conversion and Management*, *75*, 311–318.

Dao, L. A., Dehghani-Pilehvarani, A., Markou, A., & Ferrarini, L. (2019). A hierarchical distributed predictive control approach for microgrids energy management. *Sustainable Cities and Society*, *48*.

Dao, L. A., Piroddi, L., & Ferrarini, L. (2015). Impact of wind power prediction quality on the optimal control of microgrids. *5th International Conference on Clean Electrical Power: Renewable Energy Resources Impact, ICCEP 2015*.

Dao, Le Anh, Piroddi, L., & Ferrarini, L. (2017). Ensemble Methods for PV Power Production Prediction. *Internaltional Conference on Clean Electrical Power (ICCEP)*, 184–189.

Hales, D. (2018). *Renewables 2018 Global status report*.

Hossain, M., Mekhilef, S., Danesh, M., Olatomiwa, L., & Shamshirband, S. (2018). Application of extreme learning machine for short term output power forecasting of three grid-connected PV systems. *Journal of Cleaner Production*, *167*, 395–405.

Ji, W., & Chee, K. C. (2011). Prediction of hourly solar radiation using a novel hybrid model of ARMA and TDNN. *Solar Energy*, *85*(5), 808–817.

Morari, M., & H. Lee, J. (1999). Model predictive control: Past, present and future. *Computers and Chemical Engineering*.

Schölkopf, B., & Smola, A. J. (2002). *Learning with Kernels: Support Vector Machines, Regularization, Optimization, and Beyond.* The MIT Press.

Sulaiman, S. ., Rahman, T. . A., & Musirin, I. (2009). Partial Evolutionary ANN for Output Prediction of a Grid-Connected Photovoltaic System. *International Journal of Computer and Electrical Engineering*, *1*(1), 40–45.

Teleke, S., Baran, M. E., Bhattacharya, S., & Huang, A. Q. (2010). Optimal Control of Battery Energy Storage for Wind Farm Dispatching. *IEEE Transactions on Energy Conversion*, *25*(3), 787–794.

Williams, B. M. (2001). Multivariate vehicular traffic flow prediction: Evaluation of ARIMAX modeling. *Transportation Research Record*.

Yang, D., Jirutitijaroen, P., & Walsh, W. M. (2012). Hourly solar irradiance time series forecasting using cloud cover index. *Solar Energy*, *86*(12),

Zhou, Z.-H. (2012). *Ensemble Methods: Foundations and Algorithms*. Chapman and Hall/CRC.